\documentclass[draft,12pt]{article}
\usepackage{amsmath,amsfonts,palatino,amsthm,epsf}
\newcommand{\beq}{\begin{equation}}
\newcommand{\eeq}{\end{equation}}

\newcommand{\f}{\begin{equation}}
\newcommand{\ff}{\end{equation}}
\newcommand{\blankline}{\vskip .3cm}

\begin{document}

\title{The case for  background independence\footnote{This is partly  based on the  text of a talk given to a meeting of the British Association for the Philosophy of Science, in July 2004, under the title "The relational idea in physics and cosmology."}}
\author{Lee Smolin\thanks{Email address:
lsmolin@perimeterinstitute.ca}\\
\\
\\
Perimeter Institute for Theoretical Physics,\\
35 King Street North, Waterloo, Ontario N2J 2W9, Canada, and \\
Department of Physics, University of Waterloo,\\
Waterloo, Ontario N2L 3G1, Canada\\}
\date{\today}
\maketitle

\begin{abstract}

The aim of this paper is to explain carefully the arguments behind the assertion that the 
correct quantum
theory of gravity must be background independent.  We begin by recounting how the 
debate over whether quantum gravity must be  background  independent  is a continuation of a long-standing argument in the history of physics and philosophy over whether space and time are relational or absolute. This leads to a careful statement of what physicists mean when 
we speak of background independence. Given this we can  characterize 
the precise sense in which general relativity is a background 
independent theory.  The leading background independent 
approaches to quantum gravity are then discussed, 
including causal set models, loop quantum gravity 
and dynamical triangulations and their main achievements 
are summarized along with the problems that remain open.   
Some first attempts to cast string/$\cal M$ theory into a 
background independent formulation are also mentioned.   

The relational/absolute debate has implications also for other issues such as   
unification and how the parameters of the standard models
of physics and cosmology are to be explained.   
The recent issues concerning the string theory landscape are reviewed  
and it is argued that they can only be resolved within the context of a
background independent formulation. 
Finally, we  review
some recent proposals to make quantum theory more 
relational.    
\end{abstract}

\newpage
\tableofcontents
\newpage

\section{Introduction}

During the last three decades research in theoretical physics has focused on four key problems, which, however,   remain unsolved.  These are

\begin{enumerate}

\item{} The problem of quantum gravity.

\item{} The problem of further unifying the different forces and particles, beyond the partial 
unification of the standard model.

\item{} The problem of explaining how the parameters of the standard models of particles physics and cosmology, including the cosmological constant, were chosen by nature.

\item{} The problem of what constitutes the dark matter and energy, or whether the evidence for them are to be explained by modifications in the laws of physics at very large scales. 

\end{enumerate}

One can also mention a fifth unsolved problem, that of resolving the controversies concerning the foundations of quantum
mechanics. 

All five problems have remained unsolved, despite decades of determined effort by thousands of extremely talented people.   While a number of approaches have been studied, most expectations have been put on string theory as it appears to provide a uniquely compelling unification of physics.  Given that  the correct perturbative dynamics for gauge fields, fermions and gravitons emerges from a
simple action expressed in terms of worldsheets and that, in addition,  there are strong indications that the quantum corrections to these processes are finite to each order of string perturbation theory\cite{finite}, it is hard
not to take string theory seriously as a hypothesis about the next step in the unification
of physics.   At the same time, there remain open problems.  

Despite knowing a great deal about the different perturbative string theories and the dualities that relate them, it is widely believed that a more fundamental formulation exists.  This would give us a set of equations,  solutions to which would give rise to the different perturbative string theories.  While there is a lot of evidence for the existence of this more fundamental formulation, in the dualities that relate the different string perturbation theories, there is as yet no agreed upon proposal as to either the principles or the equations of this formulation. 

It is also unfortunately the case that the  theory makes, as yet,  no falsifiable predictions for doable experiments, by which the applicability of the theory to nature could be checked. This is because
of the landscape of discrete vacau which have been uncovered in the last few years.  
Powerful effective field theory arguments have made it plausible that  the theory comes in  an infinite number   of versions\cite{iinfinite-versions}.  These  appear to correspond to an infinite  number of possible universes and low energy phenomenologies.  Even if one imposes the minimal phenomenological
constraints of a positive vacuum energy
and  broken supersymmetry, there are argued to be 
still a vast ($>10^{300}$) number of theories\cite{KKLT}. 
There thus appears to be no uniqueness and no predictability so far as observable parameters are concerned, for example, one can get any gauge group and many different spectra of Higgs and fermions.

Of course, these two issues are related.  It seems very likely that the challenge posed by the landscape would be resolved if we had a more fundamental formulation of string theory.  This would enable us to establish which of the vacua described by effective field theory are truly solutions to the exact theory.  It would also allow us to study the dynamics of transformations between different vacua.  

Another striking feature of the present situation is that  we  have no unique predictions for the post-standard model physics which will be explored in upcoming experiments at the LHC.  This is true in spite of the fact that we have had three decades since the formulation of the standard model to discover a convincing theory that would give us unique predictions for these experiments. The theory many of our colleagues believe, the supersymmetric extension of the standard model, has too many parameters to yield  unique predictions for those experiments.   

It is beyond doubt that research in string theory has nonetheless led to a large number of impressive results and conjectures, some of great mathematical beauty.  Several mathematical conjectures have been suggested by work in string theory, that turned out to be provable by more rigorous means.  A number of interesting conjectures and results have been found for the behavior of supersymmetric gauge theories.  All of this suggests that string theory has been worth pursuing.   At the same time, the present situation is very far from what was expected when people enthusiastically embraced string theory 20 years ago. 

If so much effort has not been rewarded with success, it might be reasonable to ask
whether some wrong assumption was made somewhere in the course of the development of the theory.  The purpose of this paper is to propose such an hypothesis.  This hypothesis is made with an open minded spirit, with the hopes of
stimulating discussion.

To motivate my hypothesis, we can start by observing that theorists' choices of how to approach the key issues in fundamental physics is largely determined by their views on three  crucial questions.

\begin{itemize}

\item{}{\it Must  a quantum theory of gravity must be  background independent, or can  there can be a sensible and successful background dependent approach? }

\item{}{\it How are the parameters of the standard models of physics and cosmology  to be determined?}

\item{}{\it Can a cosmological theory be formulated in the same language we use for descriptions of
subsystems of the universe, or does the extension of physics from local to cosmological require new principles or a new formulation of quantum theory?}

\end{itemize}

It is the first issue that divides most string theorists from those who pursue alternative
approaches to quantum gravity.  

The second issue determines the attitude different people take to  the landscape.  There are,
roughly, three possible approaches:   1) a unique theory leading
to unique predictions. 2) Anthropic approaches, according to which our
universe may be very different from a typical member of an ensemble or landscape of theories\footnote{A critique of
the attempts to resolve the landscape problem through the anthropic principle is given in
\cite{me-anthropic}.}.  3)  Dynamical, or 
evolutionary approaches, according to which the dynamics of reproduction of universes 
results in our universe being a typical member of the ensemble\cite{lotc}. 
The first has been, traditionally, 
the basis of the hopes for a unified theory, but the recent
results suggest that unification leads not to a single, unique theory, but a 
multitude of possible theories. This leaves the other two options.  

The third issue has been long appreciated by those who have attempted to formulate a sensible
quantum theory of cosmology,  but it recently has been raised in the contexts of attempts to
resolve the problems of the landscape in terms of cosmological theories and hypotheses.  

In this paper I would like to make two  observations and an hypothesis about these issues.

\blankline
\blankline

\begin{itemize}

\item{}These three debates are closely related and they are unlikely to be resolved
separately. 

\item{}  These three debates are aspects of a much older debate, which has been central to 
thinking about the nature of space and time going back to the beginning of physics. This is the {\it debate between relational and absolute theories of space and time.}

\end{itemize}

In particular, as I will explain below, background dependent attempts at quantum gravity and anthropic approaches to the landscape are the contemporary manifestations of the absolute side of the old debate. 
Similarly,  background independent approaches to quantum gravity and dynamical or evolutionary approaches to the landscape are firmly within the relational tradition.  

Now here is my thesis, which it is the task of this essay to support:  

\blankline
\blankline

{\it The reason that we do not have a fundamental formulation of string theory, from which it might be possible to resolve the challenge posed by the landscape, is that  it has been so far developed as a background dependent theory. This is despite there being compelling arguments that a fundamental theory must be background independent.   Whether string
theory turns out to describe nature or not, there are now few alternatives but to approach the
problems of unification and quantum gravity from a background independent perspective.}  

\blankline
\blankline

This essay is written with the hope that perhaps some who  have avoided thinking about background independent theories might consider doing so now.   
To aid those who might be so inclined, in the next section I give a sketch of how the absolute/relational debate has shaped the history of physics since before the time of Newton. Then, in section 3, I explain precisely what is meant by relational and absolute theories.  Section 4 asks whether general relativity is a relational theory and explains why the answer is: partly. We then describe, in section 5, several relational approaches to quantum gravity.  There have been some remarkable successes, which show that it is possible to get highly non-trivial results from background independent approaches to
quantum gravity\cite{invitation}. At the same time, there remain open problems and challenges. 
Both the successes and open problems yield lessons for any future attempt to make
a background independent formulation of string theory or any other quantum theory of gravity. 

Sections 6 to 8 discuss what relationalism has to offer for the problems in particle physics such as unification and predictability.  It is argued that the apparent lack of predictability 
emerging from studies of the string theory landscape is a symptom of relying on 
background dependent methodologies in a regime where they cannot offer sensible
answers.  To support this, I show that 
relationalism suggests methodologies by which multiverse theories may nevertheless make falsifiable predictions.  

Many theorists have asserted that no approach to quantum gravity should be taken
seriously if it does not offer a solution to the cosmological constant problem. In section 
9 I show that relational theories do offer new possibilities for how that most recalcitrant
of issues may be resolved. 

Section 10 explores another application of relationalism, which is to the problem of how to extend   quantum theory to cosmology.   I  review several approaches which have been called "relational quantum theory."  These 
lead to formulations of the holographic principle suitable for quantum gravity and cosmology.

\blankline
\blankline

\section{A brief history of relational time}

The debate about whether space and time are relational is central to the history of physics.   
Here is a cartoon sketch of the story\footnote{A full historical treatment of the relational/absolute
debate is in Barbour's book, \cite{julian-first}}.  

Debate about the meaning of motion go  back to the Greeks. But the issues of interest for us
came into focus  when Newton proposed his
form of dynamics in his book {\it Principia Mathematica}, published in 1687.    Several of his 
rough contemporaries, 
such as Descartes, Huygens and Leibniz espoused 
{\it relational} notions of space and time, according to 
which space and time are to be defined only in terms 
of relationships among real objects or
events.   Newton broke with his contemporaries to espouse  
an {\it absolute}  notion of space and time, according to which  
the geometry of space and time provided a fixed, immutable and 
eternal background, with respect to which particles moved.  
 Leibniz responded by proposing arguments for a relational view 
that remain influential to this day\footnote{Some essential texts, accessible to 
physicists, are \cite{leibniz}.}.  

Leibniz's argument for relationalism was based on two principles, which have been the focus of many  books and papers by philosophers to the present day. 
The {\it principle of sufficient reason} states that it must be possible to give a rational justification for every
choice made in the description of nature.  I will refer the interested reader to the
original texts\cite{leibniz} for the arguments given for it, but it is not hard to
see the relevance of this principle for contemporary theoretical physics. 
A theory that begins with the
choice of a background geometry, among many  equally consistent choices, violates this principle.  So does a theory that allows some parameters to be
freely specified, and allows no mechanism or rational argument why one value
is observed in nature.  

One circumstance that the principle of sufficient reason may be applied to is spacetimes with
global symmetries.  Most distributions of matter in such a space  will not be invariant under
the symmetries.  One can then always ask, why is the universe where it is, rather than
ten feet to the left, or rotated 30 degrees?  Or, why did the universe not start five minutes
later? This is sometimes called the problem of under determination: nothing in the laws
of physics answers the question of why the whole universe is where it is, rather than translated
or rotated. 

As there can be no rational answer why the universe is where it is, and not
ten feet to the left, the principle of sufficient reason says this question should not
arise in the right theory.  One response is to demand a better theory in which there
is no background spacetime.  If all there is to space is an emergent description of relations
between particles,  questions about whether the whole universe can
be translated in space or time cannot arise.  Hence, the principle of sufficient reason 
motivates us to eliminate fixed background spacetimes from the formulation of physical law.

Conversely, if one believes that the geometry of space is going to have an absolute character,  fixed in advance, by some a priori principles,  you are going to be led to posit a homogeneous geometry.
For what, other than particular states of matter, would be responsible for inhomogeneities in the
geometry of space?  But then spacetime will have symmetries which leave you prey to the
argument just given.  So from the other side also, we see that the principle of sufficient
reason is hard to square with any idea that spacetime has a fixed, absolute character.  

One way to formulate the argument against background spacetime is through a second
principle of Leibniz,   {\it the identity of  the indiscernible}. This  states that any
two entities which share the same properties are to be identified. Leibniz argues
that were this not the case, the first principle would be violated, as there would
be a distinction between two entities in nature without a rational basis. 
If there is no experiment that could tell the difference between the state in which the universe is
here, and the state in which it is translated 10 feel to the left, they cannot be distinguished. The 
principle says that they must then be identified.  In modern terms, this is something like saying
that a cosmological theory should not have global symmetries, for they generate motions and
charges that could only be measured by an observer at infinity, who is hence not part of the universe.
In fact, when we impose the condition that the universe is spatially compact without boundary,
general relativity tells us there are no global spacetime 
symmetries and no non-zero  global conserved charges\footnote{That is, special solutions may have symmetries. But, as we will discuss in section 4,  there are no symmetries acting on the
space of physical solutions of the theory, once these have been identified with equivalence classes 
under diffeomorphisms\cite{karel}. }.

But it took physics a long time to catch up to Leibniz's thinking.  
Even if philosophers were convinced that Leibniz had the better argument, Newton's view was easier to develop, and took off, whereby Leibniz's remained philosophy. This is easy to understand:  a physics where space and time are absolute can be developed one particle at a time, while a relational view requires that the properties of any one particle are determined self-consistently by the whole universe.  

Leibniz's criticisms of Newton's physics were sharpened by several thinkers, the most influential of which was Mach\cite{Mach}, who  in the late 19th century gave an influential critique of Newtonian physics on the basis of its treatment of  acceleration as absolute. 

Einstein was among those whose thinking was changed by Mach.  There is a certain historical 
complication, because what Einstein called "Mach's principle" was not exactly what Mach wrote.
But that need not concern us here.   The key idea that Einstein got from, or read into, Mach,  was that
acceleration should be defined relative to a frame of reference that is dynamically determined by the configuration of the whole universe, rather than being fixed absolutely, as in Newton's theory.  

In Newton's mechanics, the distinction between who is accelerating and who is moving uniformly
is a property of an absolute background  spacetime geometry, that is fixed independently of the history or configuration of matter.   Mach proposed, in essence eliminating absolute space as a cause of the
distinction between accelerated and non-accelerated motion, and replacing it with a dynamically determined distinction.   This resolves the problem of under-determination, by replacing an a priori background with a dynamical mechanism.  
By doing this Mach showed us that a physics that respects Leibniz's principle of sufficient
reason is more predictive, because it replaces an arbitrary fact with a dynamically caused and  observationally
falsifiable relationship between the local inertial frames and the distribution of matter in the universe.  
This for the first time made it possible to see how, in a theory without a fixed background, properties
of local physics, thought previously to be absolute,  might be genuinely explained, self-consistently,
in terms of the whole universe. 

There is a debate about whether general relativity is "Machian", which is partly
due to confusion over exactly how the term is to be applied.  But there is no doubt that general relativity
can be  characterized as  a partly relational theory, in a precise sense that I will explain  below.  

 To one schooled in the history of the 
relational/absolute debate\footnote{The understanding that working physicists like myself have of  the  relevance of the relational/absolute debate to  the physical interpretation of general relativity and contemporary efforts towards quantum gravity is due mainly to the writings and conference talks of a few 
physicists-primarily John
Stachel\cite{stachel} and Julian  Barbour\cite{julian-relational}.  Also  important were the efforts of philosophers who, beginning in the early 90's  were kind enough to come to conferences on quantum gravity and engage us in discussion.}, it is easy to understand
the different  choices made by different theorists as reflecting different expectations and understandings of that debate\cite{more-relational}.  
The same can be said about the debates about the merits of the Anthropic Principle as a solution to the very puzzling
situation that string theory has found itself in recently\cite{me-anthropic}.  
To explain why, we need some precise definitions. 
 
\section{What physicists talk about when we talk about relational space and time}

While many physicists have been content to work with background dependent theories, from the earliest
attempts at quantum gravity there has been a community of those who shared the view that any approach
must be background independent.  Among them, there has been a fair amount of discussion and reflection concerning  the roots of the notion of background independence in older relational views of space and time.  From this has
emerged a rough consensus as to what 
may be called the physicists' relational conception of space and time\footnote{Philosophers
distinguish several versions of relationalism\cite{simon}, among which,  what is
described here is what some philosophers call {\it eliminative relationalism.}}.  

Any theory postulates that the world is made up of a very large collection of elementary entities (whether particles, fields, or events or processes.) Indeed, the fact that the world has many things in it is essential for these considerations-it means that the theory of the world may be expected to differ in important aspects from models that describe  the motion of a single particle, or a 
few particles in interaction with each other. 

The basic form of a physical theory is framed by how these many entities acquire properties.  In an absolute framework the properties of any entity are defined with respect to a single entity-which is presumed to be unchanging.  An example is the absolute space and time  of Newton, according to which  positions and motions are defined with respect to this unchanging entity.  Thus, in Newtonian physics the background is three dimensional space, and the fundamental properties are  a list  
of the positions of particles in absolute space as a function of absolute time: $x^a_i (t)$. 
Another example of an absolute background is a regular lattice, which is often used in the formulation of
quantum field theories.  Particles and fields have the property of being at different nodes in the lattice, but the lattice does not change in time. 

 The  entities that plays this role may be called the background for the description of physics.  The background consists of presumed entities that do not change in time, but which are necessary for the definition of the kinematical quantities and dynamical laws.  

The most basic statement of the relational view is that 

\blankline

{\bf R1}  \ \  {\it    There is no background.}

\blankline

How then do we understand the properties of elementary particles and fields?  The relational view presumes that

\blankline

{\bf R2} \ \ {\it  The fundamental properties of the elementary entities consist entirely in relationships between those elementary entities. }

\blankline

Dynamics is then concerned with how these relationships change in time.  

An example of a purely relational kinematics is a graph.   
The entities are the nodes. The properties are the connections between the nodes. 
The state of the system is just which nodes are connected and which are not.   The dynamics is given by a rule which changes the connectivity of the graph. 

We may summarize this as

\blankline

{\bf R3}   \  \  {\it  The relationships are not fixed, but evolve according to law.  Time is nothing but changes in the relationships, and consists of nothing but their ordering.  }

\blankline

Thus, we often take background independent and relational as synonymous.  The debate between philosophers that used to be phrased in terms of absolute vrs relational theories of space and time is continued in a debate between physicists who argue about background dependent vrs background independent theories. 

It should also be said that for physicists relationalism is a strategy. As we shall see, theories may be partly relational, i.e.. they can have varying amounts of background structure.  One can then advise that progress is achieved by adopting the 

\blankline

{\bf Relational strategy:}  \ \   {\it  Seek to make progress by identifying the background structure in our theories and removing it, replacing it with relations which evolve subject to dynamical law.}

\blankline

Mach's principle is the paradigm for this strategic view of relationalism.  As discussed above, Mach's suggestion 
was that replacing absolute space as the basis for distinguishing acceleration from uniform motion with the actual distribution of matter would result in a theory that is more explanatory, and more falsifiable.  Einstein
took up Mach's challenge, and the resulting success of general relativity can be taken to vindicate both
Mach's principle and the general strategy of making theories more relational.

\section{General relativity is a partly relational theory.}

We begin a more detailed discussion with general relativity.  As I will describe, general relativity
can be characterized as a partly relational theory. As such, it serves as a good example of the
power of the relational strategy.

There is one clarification that should be stated at the outset: the issue of whether general relativity 
is Machian or relational is only interesting if we take general relativity as a possible cosmological theory. This means that we take the spatial topology to be compact, without boundary.  In some models of subsystems of the universe, one does not do this. In these cases space has a boundary and one has to impose  conditions on the  metric and fields at the boundary. These boundary conditions become part of the background, as they indicate that there is a region of spacetime outside of the dynamical 
system which is being modeled.   

There is of course nothing wrong with modeling sub-systems of the universe with boundaries
on which we impose boundary conditions. One way to do this is to assume that the system under study is isolated, so that as one moves away from it the spacetime satisfies asymptotic conditions.  But the boundary or
asymptotic conditions can only be justified by the assertion that the system modeled is a 
subsystem of the universe.  No fundamental theory could be formulated in terms that
require the specification of boundary or asymptotic conditions because those conditions
imply that there is a part of the universe outside of the region being modeled.  Thus, one
cannot assert that a theory defined only with the presence of such conditions can be
fundamental.  

But at the same time, the fact that asymptotic  conditions can be imposed does not mean
general relativity is not fundamental, since it can also be formulated for cosmologies by
making the universe compact without boundary.  It does mean that 
 it  is only interesting to ask if general relativity is a relational theory in the cosmological case\footnote{But it is worth asking whether the fact that GR allows models with boundary conditions means that it is incomplete, as a fundamental theory.}.

General relativity is a complicated theory and there has been a lot of confusion about it. 
However, I will show now why it is considered to be mainly, but not purely, a relational theory.  One reason it is complicated is that there are several layers of structure.  

\begin{itemize}
\item{}Dimension
\item{}Topology
\item{}Differential structure
\item{}Signature 
\item{}Metric and fields
\end{itemize}

We denote a spacetime by $(M, g_{ab}, f)$, where $M$, refers to the first four properties, $g_{ab}$ is the metric and $f$  stands for all the other fields.  

It is true that in general relativity the dimension, topology, differential structure and signature are fixed.   They can be varied from model to model, but they are arbitrary and not subject to law.  These do constitute a background\footnote{ A very interesting question 
is whether the restriction to fixed dimension and topology is essential or may be eliminated by a deeper theory.}.

Then why do we say the theory is relational? Given this background, we can define an equivalence relation called a 
diffeomorphism. A  diffeomorphism $\phi$ is a smooth, invertible map 
from a manifold to itself\footnote{More generally to another manifold.}
\f
\phi  (M, g_{ab}, f) \rightarrow   (M^\prime, g^\prime_{ab}, f^\prime)
\ff
which takes a point $p$ to another point  $\phi \cdot p $, and drags the fields along with it by
\f
(\phi \cdot f)(p) = f(\phi^{-1} \cdot p )  
\ff
The diffeomorphisms of a manifold constitute a group, $Diff(M)$ , called the group of diffeomorphisms of the manifold.  The basic postulate, which makes GR a relational theory is 

\blankline
\blankline

{\bf R4}{  \it  A physical spacetimes is  defined to correspond, not to a single $(M, g_{ab}, f)$, 
but to an  equivalence class of manifolds, metrics  and fields under all actions of $Diff(M)$.  }
This equivalence class may be denoted $\{ M, g_{ab}, f\}$.

\blankline
\blankline

The important question for physics is what information is coded inside an equivalence class 
$\{ M, g_{ab}, f\}$, apart from the information that is put into the specification of $M$?    

The key point is that the points and open sets that define the manifold are not preserved 
under $Diff(M)$, because any diffeomorphism except the identity takes points to other points.  Thus, the information coded in the equivalence classes cannot be described simply as the values of fields at points. 

The answer is that 

1) Dimension and topology are coded in $\{ M, g_{ab}, f\}$  

2) Apart from those, all that there is, is a system of relationships between events.  Events are not points of a manifold, they are identifiable only by coincidences between the values of fields preserved by the actions of diffeomorphisms.  

The relations between events are of two kinds

\blankline

a) {\bf causal order}  (i.e. which events causally precede which, given by  the lightcone structure).

\blankline

b) {\bf measure}  (The spacetime volumes of sets defined by the causal order.)

\blankline

It can be shown that the information in a spacetime $\{ M, g_{ab}, f\}$ is completely characterized by the causal structure and the measure\cite{malamant}.  Intuitively, this is
because the conformal metric\footnote{Defined as the equivalence class of metrics
related by local conformal transformations 
$g_{ab} \rightarrow g^\prime_{ab} = \phi^2 g_{ab}$, where $\phi$ is a 
function.} determines, and is determined by,  the light cones and hence the causal structure.  The remaining conformal factor then determines the volume
element.  

The problem of problem of under-determination raised in section 2 is solved by the identification, in {\bf R4},  of
physical histories with equivalence classes.  For the spatially compact case, once we have moded out by the diffeomorphisms, there remain no symmetries on the space of solutions\cite{karel}.    But why should we
mod out by diffeomorphisms?  As Einstein intuited in his famous ``hole argument", and Dirac codified, one must mod out by diffeomorphisms if one is to have deterministic evolution from initial 
data\cite{stachel,more-relational,carlo-book}.  

This establishes that, apart from the specification of  topology, differential structure and dimension, general relativity is a relational physical theory.

\subsection{The problem of time and related issues}

As I emphasized at the beginning of this section, a truly fundamental theory cannot be
formulated in terms of boundaries or asymptotic conditions.   This, together with
diffeomorphism invariance, implies that the hamiltonian is a linear combination of
constraints\footnote{This is reviewed in \cite{carlo-book,present}.}.  This is no problem
for defining and solving the evolution equations, but it does lead to subtleties in the
question of what is an observable.   One important consequences  is that one cannot
define the physical observables of the theory without solving the dynamics. In other
words, as Stachel emphasizes, {\it there is no kinematics without dynamics}.  
This is because all observables are relational, in that they describe relations
between physical degrees of freedom. You cannot just ask what is happening at a manifold point,
or an event, labeled by some coordinate, and assume you are asking a physically meaningful question.  The problem is that  because of diffeomorphism invariance, points are not physically meaningful without a specification of how a point or event
is to be identified by the values of some physical degrees of freedom.
As a result, even observables that refer to local points or regions of physical 
spacetime are non-local in the sense that as functions of initial data they depend on
data in the whole initial slice.  

As a result, the physical interpretation of classical general relativity is more subtle
than is usually appreciated. In fact, most of what we think we understand naively about
how to interpret classical GR applies only to special solutions with symmetries, where we
use the symmetries to define special coordinates.  These methods do not apply to generic solutions, which have no symmetries.  It is possible to  give a physical interpretation to the generic solutions of the theory, but only by taking into account the issues raised
by the facts that all physical observables must be diffeomorphism invariant, and the
related fact that  the hamiltonian is a sum of constraints\cite{carlo-book}.  

We see here a reflection of Leibniz's principles, in that the interpretation that
must be given to generic solutions, without symmetries, is completely different from that
given to the measure zero of solutions with symmetries.  

One can actually argue something stronger\cite{present}:  Suppose that one could
transform general relativity into a form in which one expressed the dynamics directly in terms
of physical observables.  That is, observables which commute with all the constraints, but still measure local degrees of freedom.  Then the solutions with symmetries might just disappear.  This is because, being
diffeomorphism invariant, such  observables can distinguish points only by their having different values of fields.    Such observables must degenerate when one attempts to apply them to solutions with
symmetries.  Thus, expressed in terms of generic physical observables, there may be no
symmetric solutions.  If this is true this would be a direct realization of the identity of the 
 indiscernible in classical general relativity.  

Thus, even at the classical level, there is a distinction  between 
background independent and background dependent approaches to the physical interpretation.  
If  one is interested only in observables for particles moving within a given spacetime, one can use a construction that regards that spacetime as fixed.  But if one wants to discuss observables of the gravitational field itself, one cannot use background dependent methods, for those depend on fixing the gravitational degrees of freedom to one solution.  To discuss how observables
vary as we vary the solution to the Einstein equations we need
functions of the phase space variables that make sense for all solutions.
Then one must work on the full space of solutions, either in configuration space
or phase space.   

One can see this with the issue of time. If by time you mean time experienced by
observers following worldlines in a given spacetime, then we can work within that
spacetime.  For example, in a given spacetime time can be defined in terms of the causal
structure.  
But if one wants to discuss time in the context in which the gravitational
degrees of freedom are evolving, then one cannot work within a given spacetime. One constructs instead  a notion of time on the infinite dimensional phase or configuration space of the gravitational field itself. 
Thus, at the classical level, there are clear solutions to the problems of what is time and what is an observable in general relativity.

Any quantum theory of gravity must address the same issues.    Unfortunately, background dependent approaches to quantization
evade these issues, because they take for granted that one can use the special
symmetries of the non-dynamical backgrounds to define physical observables.   
To usefully address issues such as the problem of time, or the
construction of physical observables, in a context that includes the quantum
dynamics of the spacetime itself, 
one must work in a background
independent formulation. 

However, while the problem of time has been addressed in the context of background
independent approaches to quantum gravity, the problem has not been definitively solved.
The issue is controversial and there is strong disagreement among experts. Some believe the problem is solved, at least in principle, by the
application of the same insights that lead to its solution in classical general relativity\cite{carlo-book}.
Others believe that new ideas are needed\cite{present}.   While I will not dwell on it here, the reader should be aware that the problem of time is a key challenge that any complete background
independent quantum theory of gravity must solve.

 \section{Relationalism and the  search for the quantum theory of gravity}

Let us begin by noting that conventional quantum theories are background dependent theories. The background structures for a quantum theory include space and time, either Newtonian or in the case of QFT, some fixed, background spacetime.  There are additional background structures connected with quantum mechanics, such as  the inner product.  It is also significant that the background structures in quantum mechanics are connected to the background space and time.
For example, the inner product codes probability conservation, in a given
background time coordinate. 

Thus, when we attempt to unify quantum theory with general relativity we have to face the question of whether the resulting theory is to be background dependent or not.  There are two kinds of approaches, which take the two possible answers-yes and no.  These are called background independent and background dependent approaches.

Background dependent approaches study quantum theory on a background of a fixed classical spacetime.  These can be quantum theories of gravity in a limited sense in which they study the quantization of gravitational waves defined as moving (to some order of approximation) on a fixed background spacetime.  One splits the metric into two pieces
\f
g_{ab} = b_{ab} + h_{ab}
\ff
where $b_{ab}$ is the background metric, a fixed solution to the Einstein equations, and 
$h_{ab}$ is a perturbation of that solution.  In a background dependent approach one quantizes $h_{ab}$  using structures that depend on the prior specification of 
$b_{ab}$, as if $h_{ab}$ were an ordinary quantum field, or some substitute such as a string. 

Background dependent approaches include 

\begin{itemize}

\item{}perturbative quantum general relativity 

\item{}string theory.  

\end{itemize}

Perturbative quantum general relativity does not lead to a good theory, nor are the
problems cured by modifying the theory so as to add supersymmetry or other
terms to the field equations. 

It is hard to imagine a set of better motivated conjectures than those that drove interest
in string theory.  Had string theory succeeded as a background dependent theory, 
it would have served as a counter-argument
to the thesis of this essay\footnote{ A more detailed summary of the  results 
achieved in  string theory and other approaches to quantum
gravity, together with a list of problems that 
remain unsolved is given in \cite{howfar}.}.  Conversely, given that the problems
string theory faces seem deeply rooted in the structure of the theory, it may be
worthwhile to examine the alternative, which is background independent theories. 

in recent years  there has been healthy development of a number of different
background independent approaches to quantum gravity. These include,

\begin{itemize}

\item{}Causal sets

\item{}Loop quantum gravity (or spin foam models) 

\item{}Dynamical triangulations models.  

\item{}Certain approaches to non-commutative geometry\cite{alain-bi}.

\item{}A number of approaches that posit a fundamental discrete quantum
theory from which classical spacetime is conjectured to emerge at
low energies\cite{olaf-cosmo}.  

\item{}Attempts to formulate string theory as a background independent theory.

\end{itemize}

I will briefly describe  the first three. These are well enough 
understood to illustrate both the strengths of the relational view for quantum gravity and
the hard issues that any such approach must overcome.  

\subsection{The causal set theory}

To describe the causal set model we need the definition of a causal set. 

A causal set is a partially ordered set such that the intersection of the past and future of any pair of events is a finite set. 
The elements of the causal set are taken to be physical events and their partial ordering is taken to code the relation of physical causation. 

The basic premises of the causal set model are\cite{causalset}

\blankline

1) A history of the universe consists of nothing but a causal set.  That is, the fundamental events have no properties except their mutual causal relations\footnote{The events of a causal
set are sometimes called ``elements" to emphasize the principle that each corresponds
to a finite element of spacetime volume.}.   

\blankline

2) The quantum dynamics is defined by assigning to each history a complex number which is to be its quantum amplitude\footnote{In the causal set literature the dynamics is sometimes
formulated in terms of quantum measure theory, which is a variant of the consistent histories 
formulation of quantum mechanics.}.

\blankline 

The motivation for the causal set hypothesis comes from the expectation that the geometry of spacetime becomes discrete at the Planck scale. This leads one to
expect that, 
given any  classical spacetime $\{ M, g_{ab} \}$, one will be able 
to define  a causal set  
$C$ which approximates it. The precise sense in which this is possible is: 

\blankline

We say that a causal set $C$  approximates a classical spacetime, $\{ M, g_{ab}, f\}$, if, 
 to each event $e$  in $C$ there is an event $ eÕ$  in $\{ M, g_{ab}, f\}$ , such that 
1) the causal relations are preserved and 2) there is on average 1 event $eÕ$ coming 
from $C$  per Planck volume of $\{ M, g_{ab}, f\}$. 

\blankline

We note that when a causal set does approximate a classical spacetime, it does so because it is the result of a fair sampling of the relations that define the spacetime, which are the causal order and measure. 

However if the discrete quantum theory is to be more fundamental there should be a procedure to define the classical spacetime $\{ M, g_{ab}, f\}$ from some kind of classical or low energy limit of the causal set theory. This has not yet been achieved.  A main reason is the following problem, which we call the inverse problem for causal sets\cite{nonlocal}.

\blankline

{\bf The inverse problem for causal sets:}   Given a classical spacetime $\{ M, g_{ab}, f\}$ , it is easy to define a causal set $C$ which approximates it in the sense just defined.   But almost no causal set $C$  approximates a low dimensional manifold in this sense.  Moreover, we do not have a characterization, expressed only in terms of the relations in a causal set, $C$,  which would allow us to pick out those causal sets that do approximation spacetimes. We can only do this by first constructing classical spacetimes, and then extracting from them a causal set that approximates them.   Moreover no dynamical principle has been discovered which would generate causal sets $C$ that either directly approximate low dimensional classical spacetimes, or have coarse grainings or approximations that do so.  

\blankline

This is an example of a more general class of problems, which stems from the fact that combinatorially defined discrete structures are very different from continuous manifolds. 

A very general combinatorial structure is a graph.  The possibility of a 
correspondence between a graph and a smooth geometry is based on 
two definitions. 

\blankline

{\bf Definition:}  The metric on a graph  $\Gamma$  is defined by $g(j,k)$ 
 for two nodes $k$ and $j$  is the minimal number of steps to walk from $j$ to $k$ along the graph.  

\blankline

{\bf Definition:}  A graph $\Gamma$  is said to approximate a manifold and metric 
$\{ M, g_{ab} \}$ if there is an embedding of the nodes of $\Gamma$   into points of
 $\{ M, g_{ab}  \}$  such that the graph distance $g(j,k)$ is equal to the metric distance between the images of the nodes in  $\{ M, g_{ab} \}$.  

\blankline

It is easy to see that the following issue confronts us.  

\blankline
\blankline
{\bf Inverse problem for graphs.}  Given any $\{ M, g_{ab} \}$ it is easy to construct a graph $\Gamma$   that approximates it.  But, assuming only that the dimension is much less than the number of nodes,  for almost no graphs do there exist 
low dimensional smooth geometries that they approximate.  

\blankline

Because of the inverse problem, it is fair to say that  the causal set program has unfortunately so far  failed to lead to a good physical theory of quantum gravity, But it is useful to review the logic employed:

\blankline

{\bf Logic of the causal set program:  }

\begin{itemize}

\item{} GR is relational, and the fundamental relations are causal relations.

\item{} But GR is continuous and it is also non-quantum mechanical

\item{} We expect that a quantum theory of spacetime should tell us the set of physical events is discrete.

\item{}Therefore a quantum spacetime history should consist of a set of events which is a discrete causal structure. 

\item{} Moreover, the causal structure is sufficient to define the physical classical spacetime, so it should be sufficient to describe a fundamental quantum history.

\item{}But this program so far fails because of the inverse problem.  

\end{itemize}

Given the seriousness of the inverse problem, it is possible to imagine that the solution is that there are more fundamental relations, besides those of causality.
It should be said that this direction is resisted by some proponents of causal
sets, who are rather ``purist"  in their belief that the relation of causality is
sufficient to constitute all of physics.  But a possible answer to this question
is given by another program, {\it loop quantum gravity}, where 
causal relations are local changes in relational structures that describe
the quantum geometry of space\cite{fotini}.

\subsection{ Loop quantum gravity}

Loop quantum gravity was initiated in 1986 and is by now a well developed research program, with on the order of 100 practitioners. There is now a long list of results, many of them rigorous.  Here I will  briefly summarize the key results that bear on the issue of relational space and time\footnote{For details of the results,
including those  mentioned below, and references, see \cite{invitation}.
Books on loop quantum gravity include \cite{lqg-books,carlo-book} and review papers include \cite{lqg-reviews}.}.

\subsubsection{Basic results of loop quantum gravity}

Loop quantum gravity is based on the following observation, introduced by
Sen and Ashtekar for general relativity and extended to a large class of
theories including general relativity and supergravity in  spacetime dimensions
three and higher\footnote{See for example, \cite{positive} and \cite{invitation}
and references contained therein.}. 

\begin{itemize}

\item{}General relativity and supergravity, in any spacetime dimension greater than 
or equal to $2+1$,  can be rewritten as gauge theories, such that the configuration space is the space of a connection field, $A_a$,  on a spatial
manifold $\Sigma$.  The metric information is  contained in the conjugate momenta. 
The gauge symmetry includes the diffeomorphisms of a spacetime manifold,
usually taken to be $\Sigma \times R$.  
The dynamics takes a simple form that can be understood as a constrained
topological field theory. This means that the action contains one term, which
is a certain topological field theory called $BF$ theory, plus another term
which generates a quadratic constraint.  

\end{itemize}

Consider such  a classical gravitational theory, $T$, whose histories are described as diffeomorphism equivalence class of connections and fields,  $ \{M, A_a, f \} $. 
To define the action principle one must assume that the  topology, dimension and differential structure of spacelike surfaces, $\Sigma$ , are fixed.

The following results have then been proven\cite{invitation}:

\begin{enumerate} 

\item{} The quantization of $T$ results in a unique Hilbert space, $H$ of diffeomorphism invariant states. 
There is a recent uniqueness theorem\cite{unique}, which guarantees that
for dimension of $\Sigma$ two or greater, there is a unique quantization
of a gauge field such that i) the Wilson loops are represented by operators that  create normalizable states, 
ii)  its algebra with the operator that measures electric field flux is represented
faithfully and iii) the diffeomorphisms of $\Sigma$ are unitarily implemented without anomaly.  

This unique Hilbert space has a beautiful description.  There is a orthonormal basis of $H$  whose elements are in one to one correspondence with the embeddings of certain labeled graphs $\Gamma$  in $\Sigma$.    (The label set varies depending on the dimension, matter fields, and with supersymmetry.)

Because $H$ carries a unitary representation of $Diff (\Sigma )$ it is possible
rigorously to mod out by the action of diffeomorphisms and construct a 
Hilbert space, $H^{diff}$ of spatially diffeomorphism invariant states.  This has
a normalizable basis in one to one correspondence with the diffeomorphism
classes of the embeddings in $\Sigma$ of the labeled graphs.  

This is a very satisfactory description from the point of view of relationalism.  There is no more relational structure than a graph, as two
nodes are distinguished only by their pattern of connections 
to the rest of the graph.  The labels come from the theory representations
of a group or algebra $\cal A$.   The edges are labeled by representations  
of $\cal A$, which describe properties shared between the nodes they connect.  
The labels on nodes are invariants of $\cal A$, which likewise describe properties
shared by the representations on edges incident on those nodes. 

Because there is a background topology, there is additional information 
coded in how the edges of the graph knot and link each other. Given the 
choice of background topology, this information is also purely relational. 

\item{} A quantum  history  is defined by a series of local moves on graphs that take the  initial state to the final state\cite{fotini}.  The set of local moves in each history define a causal set.  

Hence, the events of the causal set arise from local changes in another 
set of relations, that which codes the quantum geometry of a 
spatial slice.  The structure that merges the relational structure of graphs
with that of causal sets is now called a {\it causal spin foam}. 

\item{} The amplitudes for local moves that follow from the quantization of the Einstein equations are known in closed form.  The sums over those amplitudes are known to be ultraviolet finite.  Similarly, the quantum Einstein equations in the Hamiltonian form have been implemented by exact operator equations on the states.  

In the case of a spin foam model for $2+1$ gravity coupled to massive particles, it has been shown in
detail that the theory can be re-summed, yielding an effective field theory on a non-commutative
spacetime\cite{laurent-etera}. This provides an explicit demonstration of how physics in classical
spacetime can emerge from a non-trivial background independent quantum theory of gravity. The 
resulting effective field theory has in addition deformed Poincare symmetry,which confirms, in this
case, the general conjecture that the low energy limit of loop quantum gravity has 
deformed Poincare symmetry\cite{invitation}.

\item{} The quantum spacetime is discrete in that each node of the graph corresponds to a finite quanta of spatial volume. The operators that correspond to volumes, areas and lengths are finite, and have discrete spectra with finite non-zero minimal values. Hence a graph with a finite number of nodes and edges defines a region of space with finite volume and area.  

\item{} There are a number of robust predictions concerning subjects like black hole entropy. Evidence has recently been found  that both cosmological and black hole singularities bounce, so the evolution of the universe continues through apparent classical cosmological singularities. 

\item{} There are explicit  constructions of semiclassical states, coarse grained
measurements of which reproduce classical geometries. Excitations of these states,
with wavelengths long in Planck units, relative to those classical geometries,  
have been shown to reproduce the physics of quantum fields and linearized gravitational waves on those backgrounds. 

\end{enumerate} 

\subsubsection{Open problems of loop quantum gravity:}

There are of course many, in spite of the fact that the theory is well defined.

\begin{itemize}

\item{}{\it Classical limit problem}: Find the   ground state of the theory and show that it is a 
semiclassical  state, excitations of which  quantum field theory and classical GR.  

\item{}{\it Do science problem}:  By studying the excitations of semiclassical states, 
make predictions for doable experiments that can test the theory up or down.

\item{}{\it Remove the remaining background dependence problem:}  The results so far defined depend on the fact that the dimension and topology of the spatial manifold, $\Sigma$, is fixed, so that the graphs are embedded in $\Sigma$. This helps by lessoning the inverse problem.   Can this be removed-and the inverse problem solved-so that all the structure that was background for previous theories, including dimension and topology, 
is explained as following from solutions to a relational theory\footnote{For more on the inverse problem
and its implications, see \cite{nonlocal}.}?  

We note that in  some formulations of spin foam models, the  dependence
on a fixed background topology is  dropped, so that the states and histories are
defined as pure combinatorial structures.  But this makes the problem of recovering 
classical general relativity from the low energy limit more complicated.  

\item{}{\it The problem of time}:  The different proposals that have been made to resolve
the problem of time in quantum gravity and cosmology can all be studied in detail in
loop quantum gravity and related cosmological models. While there are some interesting
results, the opinion of this author is that the problem remains open.

\end{itemize}

These are hard problems, and remain unsolved, but some progress is being made on all of them.  

It is important to mention that there are real possibilities for experimental tests of the theory. This is because the discrete structure of space and time implies modifications in the usual relations between energy and momenta
\f
E^2 = p^2 + m^2  + l_p E^3 + . . . 
\ff

This turns out to have implications for experiments currently underway, having to do with ultra high energy cosmic rays and gamma ray bursts, amongst others\footnote{See \cite{invitation} for a brief review of this
important subject, with additional references.}.  Loop quantum gravity appears to make predictions for these
experiments\cite{predictions}.

\subsubsection{Lessons from loop quantum gravity for the relational program}

So far as the relational/absolute debate is concerned, loop quantum gravity teaches us several lessons:

\begin{itemize}

\item{}So  long as we keep as background those aspects of space and time that are background for classical GR, (the topology, dimension and differential structure),
 we can find a quantum mechanical description of the metric and fields.  
Thus LQG is partly relational, in exactly the same way that GR is partly relational.  

\item{} Loop quantum gravity does give us a 
detailed description
of quantum spatial and spacetime geometry. 
There are many  encouraging results, such as finiteness, and the derivation of an explicit
language of states, histories, and observables for general background independent theories of
quantum gravity.  It is possible to do non-trivial computations to study the dynamics of quantum spacetime, and applications to physical problems  such as black holes and cosmology yield results that are sensible and, in some cases,  testable.  
It is very satisfying that the description of quantum geometry
and quantum histories are formulated using beautiful relational structures
such as graphs and causal sets.  

\item{}This description is flexible and can accommodate different hypotheses as to the
dimension of spacetime, matter couplings, symmetries and supersymmetries.  

\item{} There do remain hard open problems having to do with how a classical spacetime is
to emerge from a purely background independent description.  A related challenge is to
convincingly resolve the problem of time.  
Nevertheless, significant  progress is being made on these problems\cite{laurent-etera}, and it
even appears to be possible to derive predictions for experiment by expanding around certain
semiclassical states\cite{predictions}.  

\item{}The main barrier to making an entirely relational theory of quantum spacetime appears to be the inverse problem.  

\end{itemize}

\subsection{Causal dynamical triangulation models}

These are models for quantum gravity, based on a very 
simple construction\cite{dynamical}-\cite{AL3+1}. 
 A quantum spacetime is represented by a combinatorial structure, which consists of a large number $N$ of $d$ dimensional simplexes (triangles for two dimensions, tetrahedra for three etc.)  glued together to form a discrete approximation to a spacetime.  Each such discrete spacetime is given
an amplitude, which is gotten from a discrete approximation to the action for general relativity.  Additional
conditions are imposed, which guarantee that the resulting structure is the triangulation of some smooth manifold (otherwise
there is a severe inverse problem.)  For simplicity the edge lengths are taken to be all equal to a fundamental scale, which
is considered a short distance cutoff\footnote{There is a different, but related approach, called Regge calculus, in which 
the triangulations are fixed while the edge lengths are varied.}   One defines the quantum theory of gravity by a discrete
form of the sum over histories path integral, in which one sums over all such discrete quantum spacetimes, each weighed
by its amplitude. 

These models were originally studied as an approach to Euclidean quantum gravity (that is the path integral sums over spacetimes with Euclidean signature, rather than the Lorentzian signature of physical spacetime.) . In these models the topology is not
fixed, so one has a model of quantum gravity in which one can investigate the consequences of removing topology from the
background structure and making it dynamical\cite{dynamical}.  

More recently, a class of models
have been studied corresponding to Lorentzian quantum gravity. In these cases 
additional conditions are fixed, corresponding to the existence of a global time slicing, which restricts the topology to
be of the form of $\Sigma \times R$, where $\Sigma $ is a fixed 
spatial topology\cite{AL}\footnote{The condition of a fixed global
time slicing can be relaxed to some extent\cite{relaxed-global} }.

Some of the results relevant for the debate on relationalism include,

\begin{itemize}

\item{}In the Euclidean case, for spacetime dimensions $d>2$, the sum over topologies cannot be controlled.
The path integral is, depending on the parameters of the action chosen,  unstable to the formation of either an uncontrolled
spawning of ``baby universes", or to a crunch down to degenerate triangulations. Neither converges to
allow a coarse grained approximation in terms of smooth
manifolds of any dimension.  

\item{}In the Lorentzian case, when the simplices have spacetime dimension $d=2,3,4$, 
where the topology is fixed and the formation of baby universes suppressed, there is evidence
for convergence to a description of physics in manifolds   
whose macroscopic dimension i the same  as the microscopic dimension. 
For the case of $d=4$ there results are recent and highly significant\cite{AL,AL-3+1}.  In particular, 
there is now detailed numerical evidence for the emergence of $3+1$ dimensional 
classical spacetime at large distances from a background independent quantum
theory of gravity\cite{AL-3+1}.

\item{}The measure of the path integral is chosen so that each triangulation corresponds to a diffeomorphism
class $\{ M , g_{ab}, f \}$.  The physical observables such as correlation functions measured by averaging
over the triangulations correspond to diffeomorphism invariant relational observables in spacetime. 

\end{itemize}

These results are highly significant for quantum gravity.  
It follows that earlier conjectures about the possibility of defining quantum gravity through the Euclidean
path integral cannot be realized.  The sum has to be done over Lorentzian spacetimes to have a hope of
converging to physics that has a coarse grained description in smooth spacetimes.  Further, earlier conjectures
about summing over topologies in the path integral also cannot be realized.  

As far as relationalism is concerned we reach a similar conclusion to that of loop quantum gravity. There is evidence
for the existence of the quantum theory when structures including topology, dimension and signature are fixed, as part of the background structure, just as they are in classical general relativity. When this is done one has a completely relational
description of the dynamics of a discrete version of metric and fields.   Furthermore, in the context
of each research program there has recently been reported a detailed study showing of how classical spacetime emerges from an initially discrete, background independent theory.  
This is an analytic result in the case of spin foam models in $2+1$ dimensions, with 
matter\cite{laurent-etera}, and numerical
results in $3+1$ dimensions in the causal dynamical triangulations case\cite{AL,AL-3+1}.  
 This is very
encouraging, given that the problem of how classical spacetime emerges is the most
challenging problem facing background independent approaches to quantum gravity.

\subsection{Background independent approaches to string and $\cal M$ theory}

It has been often argued that string theory requires a background independent formulation. 
This is required, not just because any quantum theory of gravity must be background independent,
but because there is a need to unify all the different perturbative string theories into one theory.  
As this must combine theories defined on different backgrounds, it must not be restricted by the
choice of a particular background.  

There are some claims that string theory does not need a background independent formulation, and can be instead defined for fixed boundary or asymptotic conditions as dual to a field theory on a fixed background, as in the AdS/CFT correspondence.  To respond to this, it 
first should be emphasized that the
considerable evidence in favor of some form of an AdS/CFT correspondence falls
short of a proof of actual equivalence, which would be needed to say that a full quantum
theory of gravity, rather than just limits of correlation functions taken to the boundary, is coded in the dual conformal field theory \cite{we-malda,howfar}). 
 But even granting the full Maldacena conjecture 
 it is hard to see how a theory defined only in the presence of boundary or asymptotic conditions, as 
 interesting as that would be,  could be taken  as a candidate for a complete formulation of a  fundamental theory of spacetime.  
This is  because the boundary or asymptotic conditions can only be interpreted physically as
standing for the presence of physical degrees of freedom outside the theory.  For example, the timelike or null killing fields at the boundary stand for the reading of a clock which is not part
of the physical systems.  Such a formulation  cannot be applied to cosmological problems, 
where the problem is precisely to formulate a consistent theory of the entire universe as a closed system. General relativity with spatially compact boundary conditions is such a theory. Hence, it seems
reasonable to require that 
a quantum theory of gravity, which is supposed to reproduce general relativity, must also make sense as a theory of a whole universe, as a closed system.  

Some string theorists have also claimed that string theory does not need a background 
independent formulation, because the fact that string perturbation theory is, in principle,
defined on many different backgrounds is sufficient.  This assertion rests on exaggeration
and misunderstanding. FIrst, string perturbation theory is so far only defined on stationary
backgrounds that have timelike killing fields.  But this is a measure zero of solutions to the
Einstein equations.  It is, however, difficult to believe that a consistent string perturbation 
theory can be defined on generic solutions to the Einstein equations  because, 
in the absence of timelike killing fields, 
one cannot have spacetime supersymmetry, without which the spectrum will generally contain
a tachyon\footnote{Note that for none of the theories in the landscape is it known  how to 
construct the free string worldsheet theory.}. 

 More generally, this assertion misses completely the key point that general relativity is itself
 a background independent theory.  
 Although we sometimes use the Einstein's equations as if they were a machine for generating
solutions, within which we then study the motion of particles of fields, this way of seeing the theory is inadequate as soon as we want to ask questions about the gravitational degrees of freedom, themselves. 
Once we ask about the actual local dynamics of the gravitational field, we have to adopt the viewpoint which 
understands 
general relativity to be a background independent theory within which the geometry is completely
dynamical, on an equal footing with the other degrees of freedom.  The correct arena for this
physics is not a particular spacetime, or even the linearized perturbations of a particular spacetime. 
It is the infinite dimensional phase space of gravitational degrees of freedom. From this
viewpoint,  individual spacetimes are just trajectories in the infinite dimensional phase or configuration space;
they can play no more of a role in a quantization of spacetime than a particular classical orbit can play in 
the quantization of an electron.  

To ask for a background independent formulation of string theory is to ask only that it
conserve the fact that the dynamics of the Einstein equations does not require, indeed does not
allow, the specification of a fixed background metric.   For,  
if one means anything at all by a quantum theory of gravity, one certainly 
 means a theory by which the degrees of freedom of the classical
theory emerge from a suitable limit of a Hilbert space description. This does not commit oneself
to the belief that the elementary degrees of freedom are classical metrics or connections, nor does
it commit oneself to a belief that the correct microscopic dynamics have to do with the
Einstein equations.  But it does imply that a 
 quantum theory must have a limit in which it  reproduces the correct  formulation 
of general relativity as a dynamical system, which is to say in the background independent
language of the classical phase space.  It would seem very unlikely that such a background
independent formulation can emerge as a classical limit of a theory defined only on individual
backgrounds, which are just trajectories in the exact phase space.  

In fact, 
there have been a few attempts to develop a background independent approach to string and $\cal M$ theory\cite{bim,topologicalM}. 
These have been based on two lessons
from loop quantum gravity: i) Background independent quantum theories of gravity can be based on
matrix models, so long as their formulation depends on no background metric.  Such a model can be based on matrices valued in a group, as in certain formulations of spin foam models.  ii)  The dynamics of all known
gravitational theories can be understood by beginning with a topological field theory and then extending the
theory so as to minimally introduce local degrees of freedom.  This can be extended to supergravity, 
including the $11$ dimensional theory\cite{11d}

By combining these, a strategy was explored in which a background independent formulation of string
or $\cal M$ theory was to be made which is an extension of a matrix Chern-Simons theory\cite{bim}.  
The Chern-Simons theory provides a starting point which may be considered a membrane dynamics, but without embedding in any background manifold.  The background manifold and embedding coordinates then arise from classical
solutions to the background independent membrane model.  It was then found that background dependent
matrix models of string and $\cal M$ theory emerged by expanding around these classical solutions. 

A recent development in this direction is a proposal for how to quantize a certain reduction of $\cal M$ theory
non-perturbatively\cite{topologicalM}.

These few, preliminary, results, indicate that it is not difficult to invent and study hypotheses for background independent formulations of string theory.  

\section{Relationalism and reductionism}

I would now like to broaden the discussion by asking:  {\it Does the relational view have implications broader than the nature of 
space and time?}  I will argue that it does\footnote{The arguments of this  and the following sections
are developed from \cite{lotc}.}.  A starting point for explaining why is to begin with  
a discussion of reductionism.  

To a certain degree, reductionism is common sense. When a system has parts, it makes sense to base an understanding of it on the laws that the parts satisfy, as well as on patterns that emerge from the exchanges of energy and information among the parts.  In recent years we have learned that very complex patterns can emerge when simple laws act on the parts of a system, and this has led to the development of the study of complex systems.  These studies have shown  that there are useful principles that apply to such complex systems and these may help us to understand an array of systems from living cells to ecosystems to economic systems.   But this is not in contradiction to reductionism, it is rather a deepening of it.  

But there is a built in limit to reductionism.  If the properties of a complex system can be understood in terms of their parts, then we can keep going and understanding the parts in terms of their parts, and so on. We can keep looking at parts of parts until  we reach particles that we believe are elementary, which means they cannot be further divided into parts.  These still have properties, for example, we believe that the elementary particles have masses, positions, momenta, spin, and charges.  

When we reach this point we have to ask what methodology we can follow to 
explain the properties of the elementary particles?   As they have no parts, reductionism will not help us.  At this point we  need a new methodology. 

Most thinking about elementary particle physics has taken place in the context of quantum field theory and its descendants such as string theory, which are background dependent theories. Let us start by asking how well these background dependent theories have  done  
resolving the problem of how to attribute properties to particles thought to be elementary.  After this we will see if background independent theories can do better.  

In a background dependent theory, the properties of the elementary entities have to do with their relationships to the background.   This is clear in ordinary quantum
field theory, where we understand particles to be representations of 
the Poincare group and other externally imposed symmetries and gauge
invariances.  In these theories the particle states are labeled precisely
by how they transform under symmetries of the background.    The specification of the
gauge and symmetry groups are indeed part of the background, because they are
fixed for all time, satisfy no dynamical principles and do not evolve.

The search for an explanation of the properties of the elementary particles  within 
quantum field theory and string theory has been based on three hypotheses:
\blankline
\blankline
{\bf Unification:}\ \   {\it   All the forces and particles are different quantum states of some elementary entities.  }

\blankline
\blankline

This elementary entity was at first thought, by Einstein and his friends, to be a field, giving rise to the once maligned subject of unified field theory.  In more recent times it is thought to be a string. These are not so far apart, for a low energy approximation to a string theory is a unified field theory. So most actual calculations in string theory involve classical calculations in unified field theories that are descendants of the theories Einstein and friends such as Kaluza and Klein studied many years ago.  

But is unification enough of a criteria to pick out the right theory of nature? By itself it cannot be, for there are an infinite number of symmetry algebras which have the observed symmetries as a subalgebra.  There is however a second hypothesis which is widely believed.

\blankline
\blankline

{\bf Uniqueness: }\ \ {\it  There exists exactly one consistent unified theory of all the interactions and particles. } 

\blankline
\blankline

If this hope is realized, then it suffices to find that one unified theory.  The first fully consistent unified theory to be  found will be the only one that can be found and it  will thus have to be the true theory of nature.  It has even been said that, because of this, physics no longer needs experimental input to progress.  At the advent of string theory, this kind of talk was very common. The transition from physics as an experimental science to physics based on finding the single unified theory was even called the passage from modern to postmodern physics. 

Given that, in a background dependent theory, particles are classified by representations
of the symmetries of the vacuum, it follows that the more unified a theory is the larger the
symmetry of the background must be.  This leads to the conjecture of

\blankline
\blankline
{\bf Maximal symmetry:}\ \   {\it   The unique unified theory will have the largest possible
symmetry group consistent with the basic principles of physics, such as quantum theory
and relativity.  }

\blankline
\blankline

These three conjectures have driven much of the work in high energy physics the last three
decades. They led first to grand unified theories (large internal gauge groups), then to
higher dimensional theories (which have larger symmetry groups) and also to 
supersymmetry.  

While these conjectures come very naturally to anyone with training in elementary particle
physics, it must be emphasized that they have arisen from a methodology which is
thoroughly background dependent.   The idea that the states of a theory are classified
by representations of a symmetry group, however used to it we have become, makes no
sense apart from a theory in which there is a fixed background, given by the
spacetime geometry and the geometry of the spaces in which the fields live.   
Theories without a background, where the geometry is dynamical and time dependent,
such as general relativity, have no symmetry groups which act on the space of their solutions.  
In general relativity symmetries only arise as accidental symmetries of particular solutions,
they have no role in the formulation of the equations or space of 
solutions of the theory itself\footnote{It must be emphasized that we are talking here
of global symmetries, not gauge invariances.  Spaces of states or of solutions do not
transform under gauge transformations, they are left invariant.}.

So the methodology of looking for theories with maximal symmetry only makes sense in
a background dependent context.  Still, since an enormous amount of work has gone into
pursuit of this idea we can ask how far it actually gets us.

The most important thing to know about how this program turned out is that string theories at the background
dependent level did not turn out to be unique.  There turn out to be five string theories in
flat $10$ dimensional spacetime background. Each of them becomes an infinite number of theories when the background is taken to be a static but curved spacetime. Many of these
are spacetimes in which a certain number of dimensions remain flat while the others are
compactified.  

To preserve the notion of a unique unification it was conjectured
that there is nevertheless a unique unification of all the string theories, which has been called
$\cal M$ theory.  This was motivated by the discovery of evidence for conjectured duality 
transformations that relate states of the different string theories.  
This theory, which has so far not been constructed, is conjectured
to include all the string theories in $10$ dimensions, plus one more theory, which is
$11$ dimensional supergravity. This is the largest consistent possible supersymmetric
gravity theory and, at least at a classical level, naturally incorporates many of the
symmetries known or conjectured that act on states of the different string theories. 

The search for $\cal M$ theory has mostly followed the methodology 
which follows the three principles we mentioned.  One posits a maximal
symmetry algebra  ${\cal A}_{max}$ that contains at least the $11$ dimensional
super-poincare algebra and then tries to construct a theory based on it.  Candidates
for this symmetry algebra include the infinite dimensional 
algebra $E_{10}$\cite{E10}  and
compact 
superalgebras which have the $11$ dimensional superpoincare algebra as a subalgebra
such as $Osp(1,32)$ and $Osp(1,64)$.

However, not much work has been done on the problem of constructing $\cal M$ theory,
despite it being apparently necessary for the completion of the program of
unification through symmetry.   It is interesting to ask why this is. 

If $\cal M$ theory is to be
a unification of all the different background dependent string theories, and hence treat them all on an equal footing, it cannot be
formulated in terms of any single spacetime background.  Hence, we expect that 
$\cal M$ theory must be a background independent theory. 

 However,  background
independent theories are very different from background dependent theories, as we
have seen already in this essay.  One reason for the lack of interest in background
independent approaches to $\cal M$
theory might be  simply that is difficult for someone schooled in background dependent methods  to make the transition
to the study of background independent theories. 

But there is a better reason, which is that there is a built in contradiction
to the idea of $\cal M$ theory.  As I just emphasized, the idea that $\cal M$ theory
is based on the largest possible symmetry is one that only makes sense in a 
background dependent context. But as we have also just seen, $\cal M$ 
theory must be background independent.   

One way out is to posit that the symmetry of $\cal M$ theory, while acting formally like
a background, will not be the symmetry of any classical metric geometry. Indeed
this is true of the possibilities studied such as $E_{11}$ and $Osp(1,32)$.  Still they 
privilege those geometries whose symmetries are subalgebras of the posited fundamental
symmetry, such as the super-poincare algebra.  It thus seems hard to avoid a situation in
which the solutions which are background spacetimes of maximal symmetry will play
a privileged role in the theory.  This is unlike the case of general relativity in which
the solutions of maximal symmetry may have special properties, like being the ground
state with certain asymptotic conditions, but play no special role in the formulation of the
dynamics of the theory.  

If we are to formulate $\cal M$ theory as  a truly background independent theory, we
need a new methodology, tailored to background independent theories.  In the next
section we will begin a discussion of what a background independent approach
to unification might look like.  

Before we do, there is one more issue about background dependent theories that we
should consider. {\it 
If the hypothesis of unification is correct, then what accounts for the fact that the
observed particles and forces have the particular properties which distinguish them?}  

The
basic strategy of all modern theories of unification is to answer this
question with the mechanism of:

\blankline
\blankline
{\bf Spontaneous symmetry breaking:}\ \   {\it   The distinctions between the
observed particles and interactions result from a vacuum state of the theory
not being invariant under all the symmetries of the dynamics.  }

\blankline
\blankline

This means that the properties that distinguish the different particles and forces
from each other are due precisely to their relationship with a choice of background,
which is a vacuum  state of the theory.  If a theory can have different vacuum
states, which preserve different subgroups of the symmetries of the
dynamics, then the properties of the particles and forces will differ in each.
Hence we see explicitly that in these theories the properties of particles
are determined by their relationship to the background.  

However, notice that something new is happening here, which is quite
important for the relational/absolute debate.  The point of {\it spontaneous}
symmetry breaking is that 
the choice of background is a consequence of the dynamics and can
also reflect  the history of the system. 
Hence theories that incorporate spontaneous symmetry breaking take a step
in the direction of relational theories in which the properties of elementary
particles are determined by their relationships with a dynamically chosen
vacuum state. 

But if the choice of vacuum state is to be determined dynamically,  the 
fundamental dynamics must be formulated in a way that is
independent of a choice of background.  That is, the more spontaneous
symmetry breaking is used to explain distinctions between particles
and interactions, the more the fundamental theory must be background
independent.  

In conventional quantum field theories this is realized to some extent. But the
background of spacetime is generally not part of the dynamics.  But in string
theory the choice of solution can involve the geometry and topology of space and time.   
Hence, we arrive again at the necessity to ground string theory on a background independent 
theory.

\subsection{The challenge of the string theory landscape}

Before we turn to see how to approach the problem of unification from a background
independent theory, we should try to draw some lessons from the status of the
search based on background dependent methods. 

We can begin by asking what tools have string theorists used to study the problem
of unification? 

A principle tool invoked in much recent work in string theory is effective field theory.  
An effective field theory is a semiclassical field theory which is constructed to represent
the behavior of the excitations of a vacuum state of a more fundamental theory below some
specified energy scale. 
They  have the great advantage that one can study a theory expanded around a particular
solution, treating that solution as a fixed background. This lets us use many of the intuitions
and tools developed in the study of background dependent theories. But there are also
disadvantages to the use of effective field theory. One is that  the threshold of evidence
required to establish as likely a string background is weakened. Whereas it was
at first thought necessary to prove perturbative finiteness around the background to 
all orders, it is now thought sufficient to display a classical solution to an effective
field theory, which is some version of supergravity coupled to branes.  

But no  effective field theory can stand on its own, for these are not consistent microscopic theories. The reliability of effective field theory must always be  justified by an appeal to its derivation from that more fundamental theory.  In the applications where it was first developed, effective field theory is
derived as an approximation to a more fundamental theory. This is true in $QCD$ and the
standard model, as well as in its applications to nuclear physics and condensed
matter physics.  

We can see this
easily by considering cases in which we believe there is no good fundamental theory, such as
interacting quantum field theories in $5$ or more dimensions.  We can construct effective
field theories to our heart's content to describe the low energy physics in such contexts. 
These may be approximations to cutoff quantum field theories, for example, based on lattices.  But they are unlikely to  be approximations to any Poincare invariant theories.  This is because there is strong evidence that the only Poincare invariant quantum field theories in more
than $4$ spacetime dimensions are free. 

However, in string theory, effective field theory is being used in a context where we do not
know that there is a more fundamental theory.  That more fundamental theory, if it exists,
is the conjectured background independent unification of the different string theories.
But since we do not have this theory, either in the form of a set of equations or principles, we cannot be assured that it exists.    Hence, by relying on effective field theory
we may get ourselves in the situation in which we are studying semiclassical theories
which are not approximations to any more fundamental theory. 

But, nevertheless, if one insists on confining investigations to background dependent
methods, there is little alternative to  reliance on effective field theory.  
In the absence of a derivation from a full quantum theory, 
one can still 
 posit that the existence of a consistent effective field theory is sufficient to justify belief in a
string background, and see where this takes us. One requires  a weak form of consistency, which is that excitations of the solution, were they to exist, would be weakly coupled. 
 Not surprisingly, perhaps,  this approach leads to evidence for a landscape consisting of 
an infinite number of
discrete string backgrounds\cite{iinfinite-versions}. Even restricting the counting
to backgrounds that have positive vacuum energy and broken supersymmetry leads
to estimates of $10^{500}$ or more discrete vacua\cite{KKLT}.  

It is interesting to note that the term ``landscape" 
 implies the existence of  a function, $h$, the height, such that the
discrete vacua are at local minima of $h$.  In the recent literature,   the height $h$ is a potential or free energy.  While it is clear what is meant by this, it is perhaps worrying that the concept of
energy is problematic in a cosmological or quantum gravity context.  This is because,  once the gravitational degrees of freedom are
included,  the energy of cosmological spacetimes  is constrained to vanish.  All cosmological
solutions to diffeomorphism invariant theories have the same energy: zero.  There is in cosmology no ground state with zero energy, solutions with different potential energies are no more or less likely to
exist, they just expand at different rates.  
Even if the background
geometry is assumed fixed,  energy and free energy
are only defined on a background that has a timelike killing field.  

But what could the height be, if not energy?  The context  which inspired the original use 
of the term landscape in string theory\cite{lotc}, was mathematical models
of natural selection, in which   the height $h$ measures the fitness, which is the
number of viable\footnote{meaning they will have their own progeny} progeny of a state.  
The term was  introduced in \cite{lotc} to evoke the methodologies by 
which {\it fitness landscapes} are studied.  

However, in the recent string theory literature on landscapes, the analogy to natural
selection is not invoked.  What then is the height?  If it is energy, then that implies the
existence of a fixed background, with a timelike killing field.  But what is the background,
when the space we are considering is a space of different vacua, with different
geometries and topologies?  There seems to be a confusion in which 
reference to a structure that depends on a fixed background
is being invoked in the description of the space of possible backgrounds.  

Another way to see that the notion of a fixed background is sneaking back into the theory is
to consider the assumptions behind the probabilistic studies of the landscape.  

There are, broadly speaking, two kinds of methods that might be brought to bear to the study of
probability distributions on such landscapes of states.  One may study  distributions that
are in  {\it equilibrium, and hence static}, or one may study {\it non-equilibrium and
hence time dependent distributions.}

Almost all the recent work on probability distributions in the string theory landscape have
taken the first kind of approach.   Some, but not all, of this work evokes what we may call 

\blankline
\blankline

{\bf The anthropic hope:}{\it  There are a vast number of unified theories, and a vast number of regions of the universe where they may act. Out of all of these, there will be a very small fraction where the laws of physics allow the existence of intelligent life. We find ourselves in one of these.  Because the number of universes and theories is so vast, theory can make few prediction except those that follow from requiring our own existence.} 
\blankline
\blankline

The reliance on the anthropic principle is unfortunate, because it can be shown 
that the use of the anthropic principle cannot lead to any falsifiable predictions.
This is argued in detail in   \cite{me-anthropic}, to which the reader is referred. 
As a result, one has to suspect that a search for a unified theory of physics that in the end
invokes the anthropic principle has reached a {\it reducto ad aburdum}. Somewhere
along the line, in the search for a unified string theory, a wrong turn has been taken. 

It could be that the wrong turn is that string theory is based on physical hypotheses
that have nothing to do with nature.  But if this is not the case, some wrong direction
must have been taken in the path that led from the conjecture of a unique unification within
string theory to the present invocations of the anthropic principle.

We can see from the survey of the situation we just made that the dilemma we have
arrived at seems to involve trying to use background dependent notions, like energy,
to do physics in a setting that must be background independent. For if there is a space
of possible backgrounds, on which we are to do dynamics, it is obvious that the form of
dynamics we employ cannot make reference to any given fixed background. 
Hence, it seems reasonable to suggest that the 
wrong turn is the failure to search for a background independent foundation
for the theory.   

It is then interesting to note that  the invocation of  static probability 
distributions harkens back to the absolute perspective. To see this we can ask, what is the 
time with respect to  which the probability distribution is considered to be static? 
 It cannot be the time within a given
spacetime background, because the probability distribution lives on the space of 
possible backgrounds.  Single universes may evolve, and may come and go, but there is
hypothesized to be nevertheless a static and eternal distribution of universes 
with different properties.  It is this distribution, that exists absolutely and for all time, that
we must go for an explanation of any properties of our universe.  
Thus, at the level of multiverses, static distributions on
landscapes have more in common with Aristotle's way of thinking about cosmology
than it does with general relativity.  

Is there then an alternative methodology for treating the landscape, which would 
naturally arise from a background independent theory?  I would like to claim there is.  The
next sections are devoted to its motivation and description.

\section{A relational approach to the problems of unification and determination of the
standard model parameters}

Let us then assume we  agree on the need to formulate a unified theory in a background
independent framework. Even without having a complete formulation of this
kind in hand, it may be of interest to ask what would a background independent approach to the problem of unification 
look like?  How would it address the problems raised in the last section? To
approach  these questions  we return to  the question of {\it  how we are to 
explain the properties of the elementary particles? }

 In a relational theory, as I explained earlier, the properties of the elementary entities can have only to do with relations they have to other elementary  entities.  Let us explore the implications of this. 

The first implication is that any relational system with a large number of parts must be complex, in the sense of having no symmetries. The reason is Leibniz's principle of the identity of the indiscernible: If two entities have the same relations to the rest, they are to be identified. Each individual entity must then have a unique set of relations to the rest.  

The elementary entities in general relativity are the events.  An event is characterized by the information coming to it, from the past. We may call the information received by an event in spacetime, the {\it view} of that event.  It literally consists of what an observer at that event would  see looking out their backwards light cone.  

It follows that any two events in a spacetime must have different views.  This implies that

\begin{enumerate}
 
\item{} {\it There are no symmetries.}

\item{}{\it  The spacetime is not completely in thermal equilibrium. }

\end{enumerate}

These are in fact true of our universe.  The universe may be homogeneous above
the  enormous scale of $300 Mpc$, but on every smaller scale there is structure.
Similarly, while the microwave background is in thermal
equilibrium, numerous bodies and regions are out of equilibrium with each other. 

Julian Barbour and I call a spacetime in which the view of 
each event is distinct a {\it Leibniz spacetime}.  We note, with some wonder, that the fact that our 
universe is not completely in thermal equilibrium is due to the fact that gravitationally bound systems have negative specific heat, and therefore cannot evolve to unique equilibrium configurations.  Furthermore, gravity causes small fluctuations to grow that would otherwise be damped.  This is why the universe is filled with galaxies and stars.  Thus, gravity, which as Einstein taught us is the force that necessarily exists due to the relational character of space and time, is at the same time the agent that keeps the world out of equilibrium and causes fluctuations to grow rather than to dissipate, which is a necessary condition for it to have a completely relational description.  

There is a further consequence of taking the relational view seriously. In a relational theory, the relations that define the properties of elementary entities are not static, they evolve in time according to some law. This means that the properties by which we characterize the interaction of an elementary particle with the rest of the universe are likely to include some which are not fixed a priori by the theory, but depend on solutions to dynamical equations.  We can expect that this applies to all of the basic properties that characterize particles such as masses and charges.

\section{Relationalism and natural selection}

How far can we go to a relational explanation of the properties of the elementary
particles in the standard model?   While the anthropic principle itself is not explanatory, it is useful to go back to its starting point, which is an  apparently true observation, which we may call

\blankline
\blankline

{\bf The anthropic observation}{\it : Our universe is much more complex (in for example its astrophysics and chemistry) than most universes with the same laws but different values of the parameters of those laws (including masses, charges, etc.)  }

\blankline
\blankline

This requires explanation.  Unfortunately  no principle has been found that explains the values of the physical parameters (which can be taken to be the parameters of the standard models of particle physics and cosmology.)  Given recent progress in string theory, there is no reason to expect such a principle to exist.  Instead, as the relational argument suggests, those parameters are environmental, and can differ in different solutions of the fundamental theory. We then require a dynamical explanation for the anthropic observation.  For it to be science, the explanation must make falsifiable predictions that are testable by real experiments.  

There is only one mode of explanation I know of, developed by science, to explain why a system has parameters that lead to much more complexity than typical values of those parameters. This is natural selection.  

It may be observed that natural selection is to some extent part of the movement from absolute to relational  modes of explanation.  There are several reasons to characterize it as such.

\begin{itemize}

\item{}{\it  Natural selection follows the relational strategy.} Before it, properties that characterize species were believed to be eternal, and to have a priori explanations. These are replaced by a characterization of species that is relational and evolves in time as a result of interactions between it and other species.  

\item{}{\it The properties natural selection acts on, such as fitness, are relational quantities, in that they summarize consequences of relations between the properties of a species and other species. }

\item{}{\it These properties are not fixed in advance, they evolve lawfully.  }

\item{}{\it A relational system requires a dynamical mechanism of individuation, leading to enough complexity that each element can be individuated by its relations to the rest.}  Natural selection acts in this way, for example, it inhibits two species from occupying exactly the same niche.  By doing so it increases the complexity, measured in terms of the relations between the different species.  

\end{itemize}

This suggests the application of the mode of explanation of natural selection to cosmology.

This has been developed in \cite{lotc}, 
and it is successful in that it does lead to predictions that are falsifiable, but so far not falsified.
The idea, briefly, is the following. 

To apply natural selection to a population, there must be:

\begin{itemize}

\item{}A space of parameters for each entity, such as the genes or the phenotypes. 
¥\item{}A mechanism of reproduction.
\item{} A mechanism for those parameters to change, but slightly, from parent to child. 
\item{} Differentiation, in that reproductive success strongly depends on the parameters.

\end{itemize}

By simple statistical reasoning, the population will evolve so that it occupies places in the parameter space leading to atypically large reproductive success, compared to typical parameter values.  (Note that creatures with randomly chosen genes are dead.)

This can be applied to cosmology: 

\begin{itemize}

\item{}The space of parameters is the space of parameters of the standard models of physics and cosmology.  This is the analogue of phenotype. At a deeper level, this is to be explained by a space analogous to genotypes such as the space of possible string theories. This leads to the term Òthe string theory landscape.Ó

\item{}The mechanism of reproduction is the formation of black holes. It is long conjectured that black hole singularities ÒbounceÓ, leading to the formation of new ÒuniversesÓ through new Òbig bangs.Ó   There is increasing evidence that this is true in loop quantum gravity. 

\item{}We may conjecture that the low energy parameters do change in such a bounce. There are a few calculations 
that support this\cite{me-anthropic}.

\item{}The mechanism of differentiation is that universes with different parameters will have different numbers of black holes. 

\end{itemize}

This leads to a simple prediction: our universe has many more black holes than universes with random values of the parameters. This implies that most ways to change the parameters of the standard models of particle physics and cosmology should have fewer black holes.  

This leads to testable predictions. I'll mention one here: there can be no neutron stars with masses larger than 1.6 times the mass of the sun. I will not explain  here how this  prediction follows, but simply note that it is falsifiable\footnote{Details of the argument can be found in \cite{lotc} and\cite{me-anthropic}. }.  So far there all neutron stars observed have masses less than 1.45 solar masses, but new ones are discovered regularly.  

\section{What about the cosmological constant problem?}

It is becoming clearer and clearer that the hardest problem faced by
theoretical physics is the problem of accounting for the small value of
the cosmological constant problem.  The problem is so hard that it
constitutes the strongest arguments yet given for an anthropic 
explanation, following an argument of Weinberg\cite{weinberg-anthropic} \footnote{See \cite{me-anthropic} for a summary, references and critique.} 

Given that background dependent theories have failed to resolve it, it is
important to ask whether background dependent approaches have done any better?

We mention several interesting results  here:

\begin{itemize}

\item{}There is an argument for the relaxation of the cosmological constant
in LQG, analogous to the Pecci-Quinn mechanism\cite{steph-cc}.  This relies on a 
connection between the cosmological constant and parity breaking, which is natural 
within LQG.  

\item{}Volovich has argued, in a particular example,  that if spacetime is emergent from more fundamental quantum degrees of freedom, then there is a dynamical  mechanism which relaxes the ground state energy\cite{volovich-cosmo}. 
This mechanism is missed
if one formulates the theory in terms of an effective field theory that describes only the low energy collective excitations on a fixed
background. 

\item{}Dreyer argues that the cosmological constant problem is in fact an 
artifact of background dependent approaches\cite{olaf-cosmo}.  
He proposes that the problem
arises from the unphysical splitting of the degrees of freedom of a fundamental,
background dependent theory into a background, which has only classical
dynamics, and quantum excitations of it.  He presents an example from
condensed matter physics in which exactly this occurs. In his model, one can
calculate the ground state energy two ways: in terms of the fundamental
hamiltonian, which is a function of the elementary degrees of freedom, and
in terms of an effective low energy hamiltonian which describes collective,
emergent low energy degrees of freedom.  The zero point energy in the latter
overestimates the ground state energy computed in the fundamental theory.

\item{} The only approach to quantum gravity that predicted the correct
magnitude of the observed cosmological  constant is the causal set 
theory\cite{causal-cosmo}.  There it naturally comes out that a universe
with many events has a small cosmological constant.  Whether the mechanism
that works there extends to other background independent approaches is
an interesting open question.  

\end{itemize}

While all these results are preliminary, what is remarkable is that new 
possibilities for resolving the cosmological constant problem appear when the
problem is posed in a background independent theory.

\section{The issue of extending quantum theory to cosmology}

Let us now turn to the third issue raised in the introduction, whether a cosmological theory can be formulated in the same language as theories of small parts of the universe, or requires a new 
formulation.  As aspect of this is the problem of quantum cosmology. In recent years
new proposals to resolve this stubborn problem have been formulated in  the context of background
independent approaches to quantum gravity.

\subsection{Relational approaches to quantum cosmology}

In the last ten years several new proposal;s have been made concerning the 
foundational issues in quantum cosmology, which have gone under the name of
 relational quantum theories\footnote{For references for this section,
see the corresponding discussion and references in \cite{invitation}.}.  These 
have been inspired by the general philosophy of relationalism. 

These approaches have been put forward, in slightly different ways, by Crane, Rovelli and Markopoulou
\cite{louis-holo,carlo-relational,algebraic-fotini}. The mathematical apparatus needed to formalize this view has been studied by Butterfield and Isham\cite{ButterfieldIsham}.  While they differ as to
details, they agree that a quantum theory of cosmology is not to be formulated in the language
of ordinary quantum mechanics. 

One way to state the problem is to ask how we understand the quantum state: Is it a complete and objective description of a physical system, in which case, how do we account for the measurement problem? Or is it a description of the information or knowledge that an observer has about a system they have isolated and studied? If this is the case, can we apply quantum theory to cosmology-or indeed to any system that contains it observers?

There is a hint of relationalism in Bohr, who argued for a view something like the latter. Bohr always insisted that while there must be a line between the system and observer, that line is flexible, it may be drawn anywhere.  This is frustrating for those who want to believe in a realist interpretation of the quantum state. A realist would argue that the observer and her instruments are physical systems.
Consequently there must be a description in which they are included in the system being studied. Bohr replies there is no contradiction, because now we are speaking of the knowledge a second observer has of a system containing the first observer.  According to Bohr then, each observer has a different
wavefunction, that describes the system they observe. 

Relational approaches to quantum theory formalize this point of view.  Rather than taking the Everett/many worlds view, and describing many universe in terms of a single quantum state, they posit that it requires many quantum states to describe a single universe. Each of these  quantum states corresponds to a  way of dividing the universe into two subsystems, such that one includes an observer.  

A relational approach to quantum theory was proposed by
Crane\cite{louis-holo}, in a paper that anticipated some aspects of the holographic principle\cite{holo}.  In that paper, Crane proposed that there is no quantum state associated
with the universe as a whole. Instead, there is a quantum state associated with
every way of introducing an imaginary spatial boundary, splitting the universe into two. By analogy with topological field theory, he proposed that the Hilbert
spaces on boundaries of $3+1$ dimensional spacetime should be built up
out of state spaces of Chern-Simons theory. When fully developed, this proposal
became the very fruitful isolated horizon approach to the quantum geometry
and entropy of horizons.  

Rovelli then developed a general framework for 
relational quantum theory\cite{carlo-relational}.   The approaches of Rovelli, however, left open the precise structure that is to tie together
the network of Hilbert spaces and algebras necessary to describe a whole
universe.   A template for such structure was given in the work of Butterfield
and Isham, who showed now the consistent histories formulation could
be interpreted in terms of a sheaf of Hilbert spaces\cite{ButterfieldIsham}.

Markopoulou proposed that the structure tying together the different
Hilbert spaces is the causal structure of spacetime\cite{algebraic-fotini}. 
In this formulation there is a Hilbert space for every event in a quantum spacetime.  The state at each event is a density matrix that describes the quantum information available to an observer at that event.  There are consistency conditions that proscribe how the flow of quantum information in a spacetime follows the causal structure of that spacetime.  This is a generalization of quantum theory, for there need not be a quantum state associated with the whole system. (Indeed, it is related to a large class of such generalizations studied by Butterfield and Isham.) 

This leads to a relational formulation of the holographic principle, sketched in
\cite{bi-holo}.   The basic idea is that the events are associated with
elements of surface. Each corresponds to a quantum channel, by which
information flows through from its causal past to its causal future.  The
area of such a channel is {\it defined} to be a measure of its channel
capacity.  

\subsection{Relational approaches to going beyond quantum theory}

Relational quantum theory gets us out of the paradoxes that arise from
trying to describe the universe with a single quantum state. Still, there is, unfortunately, 
a problem with these approaches.   This stems from the fact that the  system of quantum states  depends on the causal structure of spacetime being fixed.  But in a quantum theory of gravity one is supposed to take a quantum sum over all possible histories of the universe, each with a different causal structure.  This is to say that relational quantum theories appear to be as background dependent as ordinary quantum theory, it is just that they differ in how they are background dependent.  

Can there be a fully background independent approach to quantum theory?  I believe that the answer is only if we are willing to go beyond quantum theory, to a hidden variables theory.   I would
like in closing then to briefly mention work in progress in this direction. 

We know from the experimental disproof of the Bell inequalities that any viable hidden variables theory must be non-local. This suggests the possibility that the hidden variables are relational. That is, rather than giving a more detailed description of the state of an electron, relative to a background, the hidden variables may give a description of relations between that electron and the others in the universe. 

The possibility of a relational hidden variables theory is suggested by a simple counting argument:  In classical mechanics of N point particles, in 3 dimensional space there are 6N phase space degrees of freedom.  In quantum theory this is described by a complex function on the 3N dimensional configuration space-the wavefunction.  

But a relational theory has in principle $N^2$ degrees of freedom, at least one for every pair of particles. Most of these are unobservable, by any local observer, because they involve relations between particles near to us and those very far away. Thus, any working out of a relational theory will have to treat them probabilistically.  This will require a probability distribution, which is a real function on $N^2$ variables. 

A real function on $N^2$ variables has much more information in it than a complex function on $3N$ variables.  Thus, one can imagine deriving quantum mechanics for $3N$ variables from statistical mechanics for $N^2$  variables.  Such a theory would be a non-local hidden variables theory. 

This leads to a simple conjecture 

{\it Perhaps all the extra information, $N^2$  as compared to $N$, necessary for a completely relational theory, are the non-local hidden variables? }

In the last few years two such relational hidden variables theories have been written down.  Markopoulou and I have proposed one\cite{hiddengraphs}, and Stephen Adler\cite{adler-book}, proposes another. In our theory the non-local hidden variables are coded in a graph on $N$ nodes, which is argued to arise from the low energy limit of a relational theory like
loop quantum gravity.   

It is too soon to see if these theories will be successful.  But they offer hope that by taking relational ideas seriously may lead to a successful attack on all {\it five} of the problems mentioned in the introduction. 

\section{Conclusions}

In this talk I have described several partly relational, or background independent, theories:

\begin{itemize}

\item{}General relativity
\item{}Relational approaches to quantum gravity, including loop quantum gravity, causal set models,  causal
dynamical triangulation models and relational approaches to string/$\cal M$ theory. 

\item{}Relational  approaches to extending quantum theory to cosmology.

\end{itemize}

Each is partly successful. Several are more successful than less relational alternatives.  But none is completely successful and none is completely relational.  They are not completely relational because each still has background structure, which is non-dynamical and must be specified in advance.

However, I believe we do learn something very important from these examples:

\blankline
\blankline

{\it In several instances, the relational theory turns out to be more
predictive, and more falsifiable than background dependent
theories.}

\blankline
\blankline

In particular, cosmological natural selection leads to falsiifiable predictions, which anthropic
approaches to the landscape so far do not.  
Furthermore, there is the very real possibility that the Planck scale will be probed in upcoming experiments, such 
as GLAST and AUGER\cite{invitation}.  Background independent theories appear to give predictions for
these experiments\cite{predictions}.  String theory cannot, because it takes the symmetry of the background
as input.  

Why is this the case?  I only can make some brief remarks here. The difference
between relational and non-relational theories is between:

\blankline
\blankline

1)  Explanations that refer ultimately to a network of relationships amongst equally physical entities, which evolve dynamically. 

\blankline
\blankline
vrs
\blankline
\blankline

2) Explanations that refer to relationships between dynamical
entities and an a priori, non-dynamical, background.

\blankline
\blankline

The former are more constrained, hence harder to construct.  More of what is observed is subject to law, as there is no background to be freely chosen.   Hence, 
it appears that relational,  background independent theories are  more testable, and more explanatory.  

This is the reason for my provocative hypothesis.  If it is true than {\it the reason that string theory finds itself
in the situation described in the introduction is that  no background dependent theory could successfully solve the five key problems mentioned there.  If this is true, then the only thing to do is to go back and work on the
less studied road of relational theories. }

At the same time, I have tried here to explain the key problems still faced by the relational road. 
Some of these have to do with the problem of time. Others  have to do with 
the inverse problem. We saw it in the discussion of causal set models, which are the only purely relational theories I discussed.   The inverse problem  is that there are many more discrete relational structures than those that approximate local, continuous structures such as classical spacetimes.  So a purely relational theory that explains the fact that the world, at least on scales larger than the Planck scale, appears to be continuous and low dimensional, must explain why those local and low dimensional structures dominate in an ensemble of histories most of whom donÕt remotely resemble local, low dimensional structures.  

Let me close by recalling the extent to which  the last three decades of theoretical physics are anomalous, compared with
the previous history of physics.  Many ideas have been studied, but few have been subject to the only kind of test that really matters, which is experiment.   The hope behind this paper and the work it
represents is that by following the relational strategy we may be led to invent theories that are more falsifiable, whose
study will lead us back to the normal practice of science where theory and experiment evolve hand in hand. 

\section*{ACKNOWLEDGEMENTS}

My awareness of the centrality of the problem of background independence and its roots in the history of physics is due primarily to Julian Barbour, who has served as a philosophical guru to a whole generation of workers in quantum gravity.  My understanding of these issues has been deepened through conversations and collaborations with many people including Abhay Ashtekar, John Baez, Louis Crane, Chris Isham, Fotini Markopoulou, Roger Penrose, Carlo Rovelli and John Stachel. The work described in the closing section is joint work with Fotini Markopoulou and is largely based on her ideas.  Finally, many thanks to Freddie Cachazo, David Rideout and Simon Saunders for very helpful suggestions which  improved the manuscript.

\end{document}